\documentclass[journal]{IEEEtran}
\usepackage{tikz}

\usetikzlibrary{arrows.meta}
\definecolor{idblue}{RGB}{40,90,140}
\definecolor{idred}{RGB}{178,48,40}

\usepackage{amssymb}
\usepackage{booktabs}
\usepackage{amsmath}
\usepackage{graphicx}
\usepackage{chngcntr}
\usepackage{titlesec}
\usepackage{xcolor}

\newcommand{\worse}[1]{\textcolor{red}{\(\uparrow\)#1\%}}
\newcommand{\better}[1]{\textcolor{green!50!black}{\(\downarrow\)#1\%}}
\newcommand{\same}{0.00\%}

\ifCLASSINFOpdf
\else
\fi
\begin{document}
\setlength{\textfloatsep}{6pt plus 2pt minus 2pt}
\setlength{\dbltextfloatsep}{6pt plus 2pt minus 2pt}
\setlength{\floatsep}{6pt plus 2pt minus 2pt}
\setlength{\dblfloatsep}{6pt plus 2pt minus 2pt}
\setlength{\abovecaptionskip}{2pt}
\setlength{\belowcaptionskip}{0pt}

\setlength{\abovedisplayskip}{2pt plus 1pt minus 1pt}
\setlength{\belowdisplayskip}{2pt plus 1pt minus 1pt}
\setlength{\abovedisplayshortskip}{0pt plus 1pt}
\setlength{\belowdisplayshortskip}{1pt plus 1pt minus 1pt}

\renewcommand{\arraystretch}{0.92}
%
\title{On BESS-Backed Trading on the Continuous Intraday Electricity Market}
%
%
%

\author{Leo Semmelmann, Runyao Yu, Joseph Cary, and Derek Bunn
\thanks{Corresponding author: Leo Semmelmann. Leo Semmelmann is with the
Karlsruhe Institute of Technology, 76133 Karlsruhe, Germany
(e-mail: leo.semmelmann@kit.edu).
Runyao Yu is with the Delft University of Technology,
2628 CD Delft, The Netherlands; the Austrian Institute of Technology,
Giefinggasse 4, 1210 Vienna, Austria; and London Business School,
London NW1 4SA, U.K.
Joseph Cary is with the Department of Engineering Science,
University of Oxford, Oxford OX1 3PJ, U.K.
Derek Bunn is with London Business School, London NW1 4SA, U.K.}
\thanks{Manuscript submitted September 19, 2026.}}

%
%

\markboth{Preprint}%
{Semmelmann \MakeLowercase{\textit{et al.}}: On BESS-Backed Trading on the Continuous Intraday Electricity Market}
%



\maketitle

\begin{abstract}
Battery energy storage systems (BESS) on the continuous intraday market (IDC) often trade according to a rolling intrinsic algorithm, a myopic strategy which repeatedly opens and unwinds positions according to current prices throughout the trading session, before dispatching as delivery approaches. This paper introduces a novel trading strategy for storage, in which
forecast-driven round trip trading takes priority and the BESS supplies a second route
for closing out positions that the market would otherwise close at a
distressed price. The agent uses price quantile forecasts to open positions on the IDC, and closes these positions closer to delivery. The market can move against these trades, and the BESS intervenes only when a position would have to be settled at a price that the forecast predicts to be highly improbable, absorbing or serving that volume physically and restoring its state of
charge at ordinary prices afterwards. The position itself remains
loss-making; what the asset changes is the price at which it is closed, by transferring the terminal settlement price through time. Four quantile forecasting models of increasing complexity supply the thresholds, and all strategies are backtested on realised EPEX SPOT transactions for the German market area in 2024. A 40~MWh BESS in this role earns €2.06m against €1.77m for a rolling intrinsic benchmark and €1.37m for a perfect-foresight day-ahead benchmark, while consuming 167 instead of 365 available equivalent full cycles. Hence, the novel backstop strategy yields higher profits, while using the underlying BESS less, leaving capacity for other trading opportunities.

\end{abstract}


%
\IEEEpeerreviewmaketitle

\section{Introduction}
\label{introduction}

The increasing penetration of renewable energy creates challenges for power system operation, as forecast errors in renewable generation and load increase the complexity of power system scheduling~\cite{bunn2000forecasting}. Short-term electricity markets address these
challenges through day-ahead, intraday, and balancing markets. The
day-ahead market provides an important reference price for unit schedules, but forecast errors and outages can
introduce substantial deviations from prior positions~\cite{bordignon2013combining}. Balancing markets manage
real-time imbalances~\cite{ganesh2023forecasting}, while the continuous intraday market (IDC) is gaining importance across Europe for offsetting
deviations between forecast and actual generation and consumption as the share of variable renewable generation grows 
\cite{koch2019short}.

Battery energy storage systems (BESS) are well-placed to take advantage of price movements within the IDC~\cite{kittner2017energy}, particularly through temporal
arbitrage: charging during low-price periods and discharging during
high-price periods across different short-term markets. However, prior work suggests that BESS operation on wholesale markets alone does not meet investor requirements
\cite{baumgarte2020business}, and installed storage capacity continues to
lag far behind the levels projected as necessary for a renewable-based
power system~\cite{coester2020economic}. The recent surge in BESS grid connection requests suggests there is strong speculative interest in project development. However, converting this into the actual capacity growth required to meet power sector decarbonisation targets requires credible profitability assessments to avoid project attrition. Our approach provides an alternative revenue model for BESS, to help sustain investor confidence.

Our research presents a use case for BESS within an IDC trading strategy which differs markedly from current approaches within academic literature. Most research develops storage-centric trading strategies, bidding across energy and ancillary service markets while limited by the unit's technical constraints. Instead, our strategy uses the asset as a physical backstop to close out positions from a conventional
forecast-driven financial trading strategy\footnote{Financial products are conventionally cash-settled and imply no delivery. Ours are not: every position is opened on the IDC and carries a delivery obligation. By financial trading we mean an uncovered position, intended to be closed on the exchange before gate closure, as a non-physical trader would.}. A position that has moved
against the trader is loss-making either way; access to storage adds a
settlement route that does not require buying back/selling into a price spike/plunge. Our strategy outperforms a key BESS-centric approach in terms of absolute profit, while performing similarly on a profit per MWh basis. Importantly, our approach uses less than half as many BESS cycles as the BESS-centric strategy. This leaves cycles available for other bidding strategies, such as ancillary service market participation, which we do not account for in this work.

\subsection{Literature Review}

Recent studies have contributed to forecasting and modelling intraday
prices, including price formation and intertemporal dependencies
\cite{deschatre2024common, narajewski2020ensemble}, with
\cite{yu2025orderfusion} presenting a notable
end-to-end probabilistic forecasting model.However, the literature on
trading strategies for the IDC remains sparse. Existing studies
predict trading thresholds in theoretical settings
\cite{bertrand2018analysis}, forecast prices of individual products
\cite{hornek2025value}, analyse price formation for quarter-hourly products
econometrically without developing trading strategies
\cite{kremer2021econometric}, or generate prediction bands to inform
trading decisions \cite{serafin2022trading}. Where trading strategies are
proposed, as in \cite{hornek2025value}, the underlying forecasts remain
simplistic, while forecasting-focused studies instead draw on neighbouring
product data \cite{hornek2025directional} or focus on volatility
\cite{ciarreta2017modeling}.

Related work models BESS trading explicitly on the IDC using rolling intrinsic (RI) algorithms. In \cite{semmelmann2025algorithm}, the authors discretise the IDC trading problem into a step-wise optimisation under
continuous price updates and complete future price uncertainty, showing
substantial profit gains compared with a price index benchmark. Later work extends this approach to allow for re-optimisation after each change to the order book, and shows the value of high-frequency trading:
the fastest policy outperforms hourly re-optimisation by 58 percent \cite{schaurecker2026high}.
Miskiw et al. \cite{miskiw2025continuous} broaden the scope by
coordinating BESS trading across the day-ahead and continuous intraday
markets, comparing MILP, RI, and deep
reinforcement learning based bidding strategies under realistic
implementation constraints. Unlike  these strategies, our work combines a storage unit with another trading strategy. For a different goal, \cite{jeong2023deep} uses this approach to co-optimise real-time renewable energy bidding with battery control through deep reinforcement learning, using the asset to manage the imbalance exposure of a variable generation portfolio.

This literature review reveals two research gaps.
First, forecasting quality and trading strategy design are usually treated
separately, so it remains unclear how much of a forecasting advantage
actually converts into trading profit. Second,
across the studies that operate a BESS on the IDC the RI
algorithm serves as the core trading mechanism, with the BESS itself
enabling and constraining each trade through its state of charge, capacity
and cycle limits. The BESS is therefore always the subject of the
strategy. Whether a BESS can instead support a strategy that is
otherwise purely financial, and how efficiently it uses its limited cycles
in that role, has not been examined.

\subsection{Novelty and Contribution}

Our two contributions address these these gaps. First, we show that BESS can play a profitable secondary role in a financial trading strategy, inverting the usual relationship between the BESS and the
trading strategy. We use a forecast-driven financial trading
strategy as the primary mechanism and let the BESS act as a backstop that
physically settles positions which a financial strategy would otherwise settle at highly unprofitable prices. This approach is common in
operations management, where physical inventory serves as a backstop for
underperforming financial hedges in commodity procurement
\cite{kouvelis2011managing}. However, our context differs in two key respects. First, the operations management literature addresses planned
risk-pooling over a procurement horizon rather than the ad hoc resolution
of individual loss-making trades. Second, electricity cannot be economically stored in significant quantities over long time horizons in the same way as these other commodities.

Our second contribution is an evaluation framework that reflects the
conditions a trading desk actually faces. All strategies are backtested on
realised German IDC transaction data. The trader observes only the prices
of all simultaneously traded products and past transactions, and every decision uses that information alone. Every traded unit pays transaction fees and a bid-ask spread. This framework identifies the marginal value of BESS to the strategy as the difference between the same strategy operated with and without asset backing.

Applying out strategy to the German market of 2024, we show that the same 40~MWh BESS earns €2.06m against €1.77m for a rolling intrinsic benchmark on identical transaction data, while using fewer than half of its permitted cycles. Per equivalent full cycle, the backstop earns €10\,798 against €4\,844, more than twice the RI benchmark.

The remainder of this paper is structured as follows.
Section~\ref{methodology} introduces the forecasting algorithms, the
quantile-based trading strategy, the BESS backstop and the rolling
intrinsic benchmark. Section~\ref{sec:casestudy} describes the case study
for the German market in 2024. Section~\ref{sec:results} presents the
results, which we discuss further in Section~\ref{sec:discussion} and Section~\ref{sec:conclusion} concludes.

\section{Methodology}
\label{methodology}

This study introduces a novel strategy for trading BESS in IDC. Rather than building a strategy around the technical characteristics of a
BESS, we let the asset reduce the downside exposure of a financial trading
strategy on the IDC by giving loss-making positions a second settlement
route: instead of being bought back or sold off at a distressed price, they
can be resolved physically by charging or discharging the BESS. This
section introduces the forecasts, the threshold-based trading strategy,
the BESS backstop, the two benchmarks and the evaluation metrics.

\subsection{Forecasting quantiles of IDC prices}

Our trading strategy requires quantile forecasts of the ID1 price at
cycle open, one hour before delivery. We now describe our four forecasts: one benchmark that only
uses recent price history, and three machine-learning models of increasing
complexity from \cite{yu2025orderfusion}.


\textbf{Day-before benchmark:} The benchmark forecast combines the current price level with the price distribution observed one day earlier, based on \cite{semmelmann2026quantile}. Let $\pi_j$ and $q_j$ denote the price and volume of transaction $j$. The current price level of product $d$ is estimated at cycle open, $t = d - 60\,\mathrm{min}$, by the VWAP of the preceding 30 minutes, \begin{equation} \pi^{\mathrm{ref}}_d \;=\; \frac{\sum_{j \in \mathcal{W}_d} \pi_j\, q_j} {\sum_{j \in \mathcal{W}_d} q_j}, \end{equation} where $\mathcal{W}_d$ is the set of transactions of $d$ executed in $[t - 30\,\mathrm{min},\, t)$. The shape of the distribution is taken from the same delivery period one day earlier, $d^- = d - 24\,\mathrm{h}$. Let $\mathcal{T}_{d^-}$ be the set of transactions of $d^-$ executed during its active trading phase, from its cycle open to the start of its settlement phase, $[d^- - 60\,\mathrm{min},\, d^- - S)$, where $S$ denotes the lead time before delivery at which settlement begins, introduced in the following subsection. The price quantile $\alpha$ for $d^-$ is \begin{equation} \rho^{\alpha}_{d^-} \;=\; \hat{Q}^{\alpha}\!\big(\{\pi_j : j \in \mathcal{T}_{d^-}\}\big) - \pi^{\mathrm{ref}}_{d^-}, \end{equation} where $\hat{Q}^{\alpha}$ denotes the empirical $\alpha$-quantile. The forecast quantiles add these distances to the current price level: \begin{equation} Q^{\alpha}_{t,d} \;=\; \pi^{\mathrm{ref}}_d + \rho^{\alpha}_{d^-}  \end{equation} This benchmark algorithm reflects the persistence of intraday price patterns from one day to the next.

\textbf{Machine-learning forecasts:}
To investigate whether differences in forecasting performance translate into differences in trading revenues, we consider models from several model families. Linear Quantile Regression (LQR), which is widely used in electricity price forecasting, models a linear relationship between the input features and the conditional price quantiles. Its input is the latest available 15-minute VWAP at the forecast issuance time~\cite{hirsch2024simulation, yu2027orderbook}. An LSTM forecast provides a neural network-based benchmark~\cite{oksuz2019neural}. Our state-of-the-art forecast (OrderFusion~\cite{yu2025orderfusion}) incorporates order book microstructure information, which can provide additional predictive power~\cite{yu2026deeplearningelectricityprice}.

\subsection{Quantile-based trading}

We now describe the purely financial trading strategy, in which every position is opened and closed on the exchange. The underlying idea is well-established in the IDC literature: a forecast band of the price distribution separates low from high prices, so observed transactions that cross the band signal trading opportunities. Selling whenever the price exceeds such a band beats naive benchmarks on German intraday data~\cite{serafin2022trading}, and quantile forecasts used as dynamic buy and sell thresholds yield profitable strategies in a backtest on realised EPEX SPOT transactions \cite{semmelmann2026quantile}. We adopt the same principle with symmetric thresholds on both sides of the band. Since our focus is the BESS as a backstop rather than the trading rule
itself, we adopt this replicable, proven strategy from the literature;
more sophisticated rules, such as time-variant thresholds or weather signals
\cite{kuppelwieser2023intraday}, are left for future work.

\textbf{Purely financial:} We trade each quarter-hourly delivery period $d$
over its ID1 cycle, i.e.\ the final hour before delivery. The cycle opens at
$t = d - 60\,\mathrm{min}$, which coincides with the end of cross-border
trading in the shared European order book. Trading between the four
German control areas remains possible until $d - 30\,\mathrm{min}$, and
within a single control area until gate closure at $d - 5\,\mathrm{min}$.
Following the requirement to be balanced at delivery, we split the cycle
into an \emph{active trading phase} $[t,\, d - S)$ and a \emph{settlement
phase} $[d - S,\, d - 5\,\mathrm{min})$ with $S = 15\,\mathrm{min}$, and
carry no position beyond gate closure. This is in line with regulation in Germany that requires a balanced position in every
period \cite{stromnzv}, and prohibits deliberate speculation on balancing prices.

At cycle open, a forecaster supplies a quantile pair
$\big(Q^{\mathrm{lo}}_{t,d},\, Q^{\mathrm{hi}}_{t,d}\big)$ with
$Q^{\mathrm{lo}}_{t,d} \le Q^{\mathrm{hi}}_{t,d}$, which is held fixed for
the cycle. We evaluate two threshold variants: the outer band
$(Q^{10}_{t,d},\, Q^{90}_{t,d})$ (i.e. selling at the first instance when prices cross the upper band and buying back when prices cross the lower band) and the band $(Q^{50}_{t,d},\, Q^{50}_{t,d})$, in which the lower and upper thresholds coincide at the median. Trading around a single value is still well defined, because entries must clear the threshold by the round-trip cost $m$ introduced below, which separates the buy trigger $Q^{50}_{t,d} - m$ from the sell trigger $Q^{50}_{t,d} + m$ by $2m$. Exits, by contrast, trigger at the median itself. Because the BESS couples concurrently traded products, we do not evaluate
cycles in isolation but process the transactions of all products as one
stream ordered by execution time (also known as stream processing \cite{stephens1997survey}): each
transaction $j = (d_j, \tau_j, \pi_j, q_j)$, with product $d_j$,
execution time $\tau_j$, price $\pi_j$ and volume $q_j$, updates the open
positions and later the BESS exactly once, using only information
available before $\tau_j$. As in \cite{semmelmann2025algorithm}, we assume that any observed transaction could have been executed by our strategy at its recorded price and volume. We note that an implementation in practice would require working with order book data. Although our forecasts use order book data, these are an external source, and we use only transaction data in this work due to computational and dataset access costs. However, applying our approach to order book data is an important area for future research.

Each executed megawatt hour incurs a cost $c = c^{\mathrm{fee}} + s/2$,
where $c^{\mathrm{fee}}$ covers exchange and clearing fees and $s$ denotes a
representative bid-ask spread, of which each leg of a round trip pays one half. Because every
position requires two legs, a new position is only opened when the observed
price clears the forecast threshold by at least the round-trip cost $m = 2c$. The prices at which our strategies trade are realised transaction prices, so it may appear unconventional to include a bid-ask spread on top of these. However, our data does not show which side of the trade was the aggressor, and as such we cannot filter the transactions to include only those trades in the same direction as the execution direction according to the strategy. For example, a sell action should only be able to sell at sell-aggressor prices. We include the bid-ask spread term to correct for our strategy transacting at prices from the wrong side of the order book. This yields cost-adjusted \emph{entry} thresholds
$B_{t,d} = Q^{\mathrm{lo}}_{t,d} - m$ and $S_{t,d} = Q^{\mathrm{hi}}_{t,d} + m$,
while \emph{exits} trigger at the unadjusted thresholds: a position that
already exists is unwound as soon as the market reaches its target, whereas a
new position must additionally pay for its own round trip.


Aggregate traded position for a given product $\phi_d$ must fall within
$[-\bar{V}, \bar{V}]$, where $\bar{V}$ denotes the trading budget (in MWh) per product, to reflect internal risk-management limits.

The parametrisation of $c^{\mathrm{fee}}$, $s$ and $\bar{V}$ is
given in Section~\ref{sec:casestudy}. When a transaction is observed at $\pi_j \le Q^{\mathrm{lo}}_{t,d}$,
any open short position is bought back; if the price also satisfies
$\pi_j \le B_{t,d}$, the remaining volume opens a long position. When a
transaction is observed at $\pi_j \ge Q^{\mathrm{hi}}_{t,d}$, any open long
position is sold; if the price also satisfies $\pi_j \ge S_{t,d}$, the
remaining volume opens a short position. Executed volume is always capped
by the transaction volume $q_j$ and the remaining trading budget. In the \emph{long-only} mode, short positions are disallowed
($\phi_d \ge 0$), while the \emph{long/short} mode permits both sides, a
distinction common in financial markets \cite{brush1997comparisons} and
in electricity trading practice.

Close to delivery, we define a settlement period, in which we unwind the existing open position only, settling at each observed transaction price until the position is closed. The profit of product $d$ is the net
trading cash flow minus costs,
\begin{equation}
    R_d = \sum_{j \in \mathcal{S}_d} \pi_j\, v_j
        - \sum_{j \in \mathcal{B}_d} \pi_j\, v_j
        - c \sum_{j \in \mathcal{S}_d \cup \mathcal{B}_d} v_j ,
\end{equation}
where $\mathcal{B}_d$ and $\mathcal{S}_d$ denote the executed buys and
sells with executed volumes $v_j \le q_j$.

\textbf{Asset-backed:} The asset-backed variant couples the purely financial
strategy with a BESS that physically settles those positions whose financial
close-out would be exceptionally costly. The BESS is characterised by its
energy capacity $E$ (MWh), power limit $\bar{P}$ (MW), charging and
discharging efficiencies $\eta^{\mathrm{ch}}, \eta^{\mathrm{dis}}$, and state
of charge $\sigma \in [0, 1]$, with the operational constraints (SOC
dynamics, power limits, no simultaneous charging and discharging, and a
daily cycle limit expressed as an energy-throughput cap) adopted from
\cite{miskiw2025continuous}. Within one quarter-hourly delivery period the
BESS can move at most $\bar{e} = \bar{P} \Delta_d$ with $\Delta_d =
0.25$\,h. We require a duration of at least one hour ($E \ge \bar{P} \cdot
1\,\mathrm{h}$), so that the BESS can absorb up to four consecutive
delivery periods settling in the same adverse direction. Most batteries
being deployed today meet this comfortably, with newly commissioned
projects averaging around three hours of duration \cite{iea2026battery}. The strategy maintains a target state of charge $\sigma^{\star}$ that preserves this headroom: in the long/short mode $\sigma^{\star} = 0.5$, keeping room in both directions, while in the long-only mode $\sigma^{\star} = 0$, since distressed positions can only ever be \emph{absorbed} by charging, and an empty BESS maximises the absorbable volume.

The BESS is activated only when the market trades adversely relative to the opening position during the settlement phase of
a product. Let
$w_{t,d} = Q^{90}_{t,d} - Q^{10}_{t,d}$ denote the width of the forecast
interquantile band, which the forecaster supplies irrespective of
the trading variant. We define \emph{distress thresholds} by widening the
trading band by a multiple $\kappa$ of this width,
\begin{equation}
    D^{\mathrm{lo}}_{t,d} = Q^{\mathrm{lo}}_{t,d} - \kappa\, w_{t,d},
    \qquad
    D^{\mathrm{hi}}_{t,d} = Q^{\mathrm{hi}}_{t,d} + \kappa\, w_{t,d},
\end{equation}
with $\kappa$ being a widening factor whose value is set in
Section~\ref{sec:casestudy}. A settlement price is seen as distressed when it
deviates from the nearest trading threshold by more than a multiple of the
entire forecast bandwidth, prices that the strategy's own predictive
distribution deems highly improbable. The lower $\kappa$, the more often the BESS is used; its effect is
examined in a dedicated sensitivity analysis in the results section.

As long as observed
transaction prices do not cross these distress thresholds, the strategy settles exactly as
in the purely financial case. Only the volume
$v_d$ that remains open at gate closure because the market traded beyond
the distress threshold is handed to the BESS: a residual long position is
absorbed by charging $e^{\mathrm{ch}} = \min\{v_d,\, \bar{e},\,
(1-\sigma)E / \eta^{\mathrm{ch}}\}$, a residual short position is served by discharging $e^{\mathrm{dis}} = \min\{|v_d|,\, \bar{e},\,
\eta^{\mathrm{dis}} \sigma E\}$, and the physically settled volume is
delivered instead of being bought back or sold off at distressed prices. Since the residual $v_d$ is bounded by the trading budget $\bar{V}$ and the BESS is sized with $\bar{e} \ge \bar{V}$, the BESS does not constrain the position sizes of the financial strategy, but rather the trading budget sets the minimum BESS size. 


Whenever the SOC $\sigma$ deviates from its target $\sigma^{\star}$, all products still in their active phase, and all subsequently opening cycles, switch from normal trading to a one-sided \emph{recovery mode} until the target is restored. After charging (absorption of a distressed long position), recovery means \emph{releasing}: the strategy only sells, at prices $\pi_j \ge Q^{\mathrm{hi}}_{t,d}$. After discharging, recovery means \emph{refilling}: the strategy only buys at $\pi_j \le Q^{\mathrm{lo}}_{t,d}$. Recovery volumes are capped by $\bar{e}$, by the physically available headroom, and by the exact distance to the target state of charge, taking partial transaction volumes where necessary, so the SOC returns exactly to $\sigma^{\star}$. These trades are flagged as physical, exempt
from financial settlement, and executed by the BESS at delivery. We note that recovery occurs only trades at prices that satisfy the unmodified quantile
thresholds, which are exactly the transactions the purely financial strategy would
also have entered. So, the backstop never buys expensively or sells
cheaply to restore its target state. If no such prices occur, the SOC
simply remains displaced and recovery continues in later products. While recovery is active, the strategy opens no new financial positions, while positions that are already open are still closed at their exit thresholds and settled as usual.

All physical volumes count toward a daily throughput cap. Following
\cite{miskiw2025continuous}, we express the cycling constraint as a limit
on energy throughput rather than pricing wear per MWh, since a
full cycle need not occur continuously but accumulates over partial charge
and discharge events, which is precisely how manufacturer warranties
define the admissible number of cycles. Every charge adds
$\eta^{\mathrm{ch}} e^{\mathrm{ch}}$ and every discharge adds
$e^{\mathrm{dis}} / \eta^{\mathrm{dis}}$ of internal energy throughput,
and one equivalent full cycle corresponds to a throughput of $2E$, one
full charge and one full discharge. The number of equivalent full cycles
used per day is therefore the accumulated throughput divided by $2E$. The
identical accounting is applied to the rolling intrinsic benchmark. Once the daily cycle cap is exhausted, the backstop deactivates for the
remainder of that day. We consider two rules for this situation. The first halts trading, meaning that no new positions are opened until the cycle cap resets on the following day, while positions that are already open are still settled financially. The second continues purely financially, so that the strategy keeps opening positions and accepts distressed settlements again.

\subsection{Rolling intrinsic benchmark}

We benchmark the asset-backed strategy against an RI
strategy, which is a common approach for operating BESS on the
IDC in both research and practice
\cite{semmelmann2025algorithm, miskiw2025continuous, schaurecker2026high}. RI repeatedly locks in price spreads that are already visible in the
market: whenever one delivery period can be bought cheaply and another
sold expensively, the BESS commits to charging in the former and
discharging in the latter. Only currently observable prices are exploited
(hence \emph{intrinsic}), and no forecast is used, which distinguishes it
fundamentally from the quantile-based strategy.

Our implementation follows the open-source framework from
\cite{miskiw2025continuous}. The trading session is discretised into
buckets of length $b = 15$ minutes. For every bucket $k$ and every
quarter-hourly delivery period $d$, we calculate
$\mathrm{VWAP}_{k,d}$ of the transactions executed in that bucket. To ensure that the price estimate reflects a liquid market, a
bucket price is only considered valid if at least $\theta = 10$
transactions were executed for the product in that bucket. Products
without a valid bucket price are treated as not tradable in that bucket,
which represents a lack of liquidity. We note that the discretisation is a simplification of the trading reality, which is necessary to reduce the complexity and computational cost of the problem. However, although this discretisation is common in the literature, we see implementations based on the open order book (such as \cite{schaurecker2026high}) or single transactions as a valuable path for future research. 

The RI algorithm then iterates over the buckets in chronological order. In
each bucket, a mixed-integer linear program determines the most profitable
adjustment of the current position, given the valid bucket prices and the
trades already committed in earlier buckets. The optimisation maximises
trading revenue subject to the technical constraints of the BESS, which
are identical to those of the previously introduced backstop variant. These comprise the state-of-charge
dynamics with the charging and discharging efficiencies
$\eta^{\mathrm{ch}}$ and $\eta^{\mathrm{dis}}$, the power limit $\bar{P}$,
the exclusion of simultaneous charging and discharging, a state of charge
between empty and full, a terminal condition that ends every day with an
empty BESS, and the daily cycle cap expressed as an internal energy
throughput limit as introduced above. For the full formulation we refer to
\cite{semmelmann2025algorithm, miskiw2025continuous}. Because each
re-optimisation only commits adjustments that are profitable at the
currently observed prices, every locked-in spread is free of price risk at
decision time. Repeating this optimisation bucket by bucket as prices
update is what makes the approach rolling.

To keep the benchmark comparable to the trading strategies, every executed megawatt hour pays the same cost $c$ per leg, covering exchange fees, clearing fees, and half the bid-ask spread, and equivalent full cycles are counted with the identical two-sided throughput accounting as for the backstop. The settlement of trades, however, differs in one important point from the implementations in \cite{semmelmann2025algorithm, miskiw2025continuous}, which settle each decision at the bucket VWAP and assume the full volume trades at that price. For a fair comparison with the quantile strategies, which only execute against observed transactions, we instead fill the decided volumes from the individual transactions executed in the bucket. Hence, realised prices can deviate from the decision price, and the rolling intrinsic strategy can realise losses despite locking in spreads that were profitable at decision time.

RI is itself a hybrid physical/financial strategy, since it unwinds
earlier positions whenever that is profitable before committing to a
delivery profile, which makes it the natural reference for our approach
despite being myopic rather than forecast-informed. The purely financial
variant thus measures the additional contribution of the BESS, and the RI
benchmark the comparative benefit of the overall approach.

\subsection{Day-ahead auction benchmark}

Our second benchmark only trades BESS on the day-ahead auction, following the bidding formulation in \cite{miskiw2025continuous}. For every delivery day, a MILP chooses the hourly buy and sell quantities that maximise profit, subject to the same constraints as above. The optimisation runs against the realised DE-LU clearing prices from EPEX SPOT \cite{EPEXSPOTSE_2026}, so the benchmark is a perfect-foresight upper bound on day-ahead arbitrage. Despite the perfect foresight of day-ahead prices, we see this as a representative and informative benchmark, since German day-ahead prices can be forecast with comparatively small errors \cite{ziel2015forecasting}, and execution at the clearing price is guaranteed by bidding at the price limits under the pay-as-cleared mechanism \cite{miskiw2025continuous}. Any forecast-based variant can only earn less, so a comparison in which this benchmark loses is conservative. Having the day-ahead benchmark nonetheless helps to set our approach in relation with alternative markets.

\subsection{Evaluation Metrics}
\label{methodology:Metrics}

We evaluate four different categories. First, we evaluate the forecast quality of the underlying quantile predictions. Then, we evaluate the profits of the purely financial, asset-backed and rolling intrinsic strategy. Finally, we also analyse efficiency metrics, indicating how well the BESS cycles are used with the underlying strategies, as well as risk measures.

\textbf{Forecasting quality:}
We use the Average Quantile Loss (AQL), also known as the pinball loss, averaged across the forecast quantiles, to evaluate probabilistic forecasts~\cite{yu2026pricefmfoundationmodelprobabilistic}. AQL is a discrete approximation of the Continuous Ranked Probability Score (CRPS). We also use the Mean Absolute Error (MAE) for point forecast evaluation.
Lower values indicate better probabilistic and pointwise
accuracy, respectively.

\textbf{Profit:} The profit of the purely financial strategy is the sum of
the product profits $R_d$ defined above. The value of the backstop is
measured by running the purely financial variant and the asset-backed
variant on the identical transaction stream and comparing their total
profits,
\begin{equation}
    \Delta R = \sum_d R^{\mathrm{AB}}_d - \sum_d R^{\mathrm{PF}}_d .
\end{equation}

with $R^{\mathrm{AB}}_d$  being the asset-backed profit and $R^{\mathrm{PF}}_d$ the profit from purely financial trading at day $d$. Our headline result is a comparison of the benchmark RI profit with the asset-backed strategies. 

\textbf{Cycle efficiency:} Since BESS cycles are the scarce resource
that warranties restrict, we also measure and analyze the profit earned per equivalent full
cycle. For the backstop, this is the marginal value $\Delta R / N$, where
$N$ is the number of equivalent full cycles consumed over the evaluation
period, and $\Delta R$ is the difference between purely financial and asset-backed profits, because the backstop only adds value on top of the purely
financial strategy. For the RI benchmark, it is the total
profit divided by the cycles used, because there the BESS generates the
entire profit. This metric expresses how efficiently each strategy
converts BESS wear into revenue.

\textbf{Risk:} As a risk-adjusted performance measure we report the daily ratio of mean daily profit to its standard deviation, which we refer to as the profit-to-risk ratio, analogous to the Sharpe ratio. We state it at daily resolution, since scaling to a longer horizon requires independent and identically distributed profits \cite{lo2002statistics}. In addition, we note the
maximum drawdown of the cumulative profit curve, which captures the tail
risk that a variance-based measure can understate.

\section{Case study}
\label{sec:casestudy}

We use DE-LU IDC transaction data for 2024 from EPEX Spot
\cite{EPEXSPOTSE_2026}, with $S = 15\,\mathrm{min}$, so that the
settlement phase spans the final ten tradeable minutes of each product.
The data cover only EPEX SPOT, which holds the largest share of German
continuous intraday trading; volumes on other exchanges such as Nord Pool
are not observed, but their prices closely track EPEX, so we expect no
material effect on the results.

Exchange and clearing fees are set to $c^{\mathrm{fee}} =$ €0.10/MWh,
following \cite{miskiw2025continuous}. The bid-ask spread is set to
$s =$ €1/MWh, informed by the volume-weighted cost-of-round-trip of
roughly €2/MWh that \cite{kuppelwieser2021liquidity} report for 10\,MWh
orders on the German IDC; we halve their estimate,
which stems from order-book data of 2017--2018, to reflect the substantial
growth in intraday liquidity since then. This yields a per-megawatt-hour cost
of $c =$ €0.60/MWh and a required threshold clearance of
$m = 2c =$ €1.20/MWh for opening a new position. The trading budget is set
to $\bar{V} = 10$\,MWh per delivery period.

The BESS is parametrised with an energy capacity of $E = 40$\,MWh and a power
limit of $\bar{P} = 40$\,MW, representative for the scale of some projects under construction in Germany today~\cite{tauber2026wertheim}. We note that larger BESS or portfolios can influence the magnitude of the results, considering liquidity effects. This is out of the scope of our current research. Per quarter hour
the asset can therefore move $\bar{e} = 10$\,MWh, which matches the trading
budget $\bar{V}$ exactly, so that the distressed residual of any single
product can always be absorbed within the power limit, while the full
capacity covers four such quarter hours in the same adverse direction.p

The round-trip efficiency is set to 0.86, following
\cite{semmelmann2025algorithm}, and applied symmetrically as
$\eta^{\mathrm{ch}} = \eta^{\mathrm{dis}} = \sqrt{0.86} \approx 0.927$ per
side. Throughput is capped at one equivalent full cycle per day, corresponding to $2E = 80$\,MWh of internal energy throughput, following the implementation in related RI literature \cite{miskiw2025continuous, semmelmann2025algorithm}. In the long/short mode the target state of charge leaves 20\,MWh of headroom in either direction, corresponding to two consecutive full-power quarter hours per direction, while in the long-only mode the full 40\,MWh are available for absorbing long positions. The distress widening factor is set to $\kappa = 0.5$.

\section{Results}
\label{sec:results}

\textbf{Forecasting quality:} Table~\ref{tab:id-forecast-loss} shows the AQL and MAE of the four forecasting algorithms for the ID1 index, evaluated on our out-of-sample transaction data from 2024. The forecasting models are optimised by three-fold time-series cross-validation with an expanding window: each fold tests on one four-month block of 2024, validates on the four months preceding it, and trains on all data from January 2022 up to that point, so that training, validation and testing are strictly chronologically separated.
As expected, OrderFusion is the most accurate forecast on both metrics, followed by the LSTM and
the LQR, which reproduces the ordering reported in \cite{yu2025orderfusion}.
The day-before benchmark is the weakest of the four, with an AQL 32\% higher than OrderFusion.
We note that a better forecast does not necessarily translate into a higher
trading profit, which is what we evaluate next.

\begin{table}[t]
\footnotesize
\centering
\caption{Metrics for quantile forecasts in 2024, with percentage differences relative to OrderFusion.}
\label{tab:id-forecast-loss}
\begin{tabular}{lrrrr}
\toprule
Model & AQL & $\Delta $ & MAE & $\Delta$ \\
\midrule
Day-before & 9.33 & \worse{31.87} & 23.86 & \worse{19.11} \\
LQR & 8.09 & \worse{14.46} & 20.74 & \worse{3.50} \\
LSTM & 8.05 & \worse{13.79} & 21.30 & \worse{6.29} \\
OrderFusion & 7.07 & -- & 20.03 & -- \\
\bottomrule
\end{tabular}
\end{table}

\begin{figure}[t]
\centering
\includegraphics[width=\columnwidth]{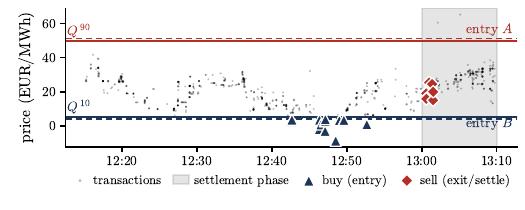}
\caption{One ID1 trading cycle of the purely financial strategy with
OrderFusion Q10/90 thresholds. Blue triangles are executed buys, and red
diamonds executed sells in the shaded settlement phase.}
\label{fig:example_cycle}
\end{figure}

\textbf{Purely financial trading:} We first illustrate how the strategy operates within a single delivery
period. Fig.~\ref{fig:example_cycle} shows one ID1 cycle of the
OrderFusion Q10/90 strategy. At cycle open the forecasting algorithm
provides the quantile band for the delivery period. Whenever a transaction
is observed below the cost-adjusted entry threshold, the strategy adds to a
long position (in this example), partly at negative prices. The position is
closed during the settlement phase at prices between €15 and 25/MWh. The
dashed lines show the effect of the entry margin (including spreads and transaction costs), because the observed
price must clear the forecast band by the round-trip cost before a position
is opened.

\begin{figure}[t]
\centering
\includegraphics[width=\columnwidth]{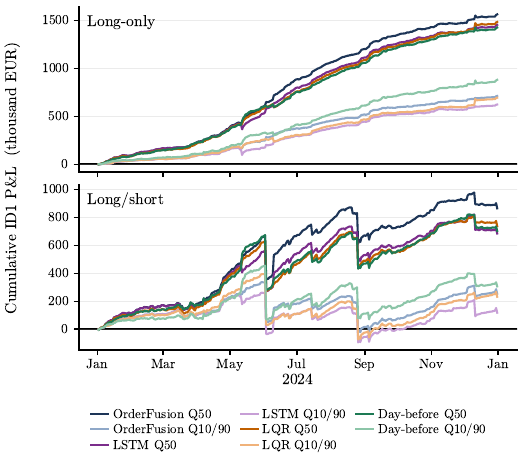}
\caption{Cumulative profit of the purely financial ID1 strategies in 2024.
Legend entries show the full-year totals in thousand EUR for the long-only
and the long/short mode.}
\label{fig:id1_modes_cumpnl}
\end{figure}

Fig.~\ref{fig:id1_modes_cumpnl} shows the cumulative profit of all purely
financial strategies over 2024 for both trading modes. We highlight three observations: First, the median strategies (Q50) earn considerably more than the outer band
strategies. Trading around the median exploits the within-session mean reversion of
intraday prices at high frequency, whereas the outer band is crossed far
less often. This is in line with \cite{kremer2021econometric}, which reports
strong mean reversion in the price formation process of the German intraday
market.

Second, the differences between the forecasting algorithms are modest in the
long-only mode, and they are much smaller than the differences in forecast
error would suggest. OrderFusion Q50 earns €1.56m and the
day-before benchmark Q50 earns €1.43m.

Third, the long/short mode earns less than the long-only mode although it
takes considerably more risk. Its cumulative profit curves fall sharply on
days with extreme price spikes, when short positions have to be bought back
at a multiple of their entry price. The profit-to-risk ratio falls from 0.703 to 0.0990.

\begin{figure}[t]
\centering
\includegraphics[width=\columnwidth]{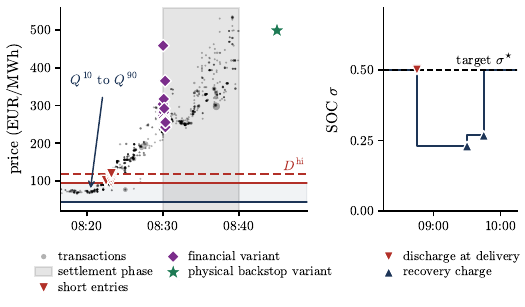}
\caption{A distressed settlement on 23 February 2024. The financial
variant buys the short position back into the spike, while the
asset-backed variant discharges the BESS and recovers afterwards.}
\label{fig:example_backstop}
\end{figure}

\textbf{Asset-backed trading and comparison with the rolling
intrinsic benchmark:} We again begin with an illustration of an example trading cycle.
Fig.~\ref{fig:example_backstop} shows a distressed settlement and how the
backstop resolves it. Prices for the 08:45 product on 23 February 2024 stay
near €80/MWh for most of the trading hour. When they cross the upper
entry threshold, the strategy opens a short position of 10~MWh at a
VWAP of €110.73/MWh. Prices then rise far beyond the distress threshold. The purely
financial strategy buys the position back during the settlement phase at a
VWAP of €269.42/MWh, so the trading session closes with a loss of €1\,587. The asset-backed strategy discharges the BESS instead and
delivers the energy physically.

The lower panel shows the consequence for the BESS. Its stored energy (denoted as SOC) drops when the physically settled volume is delivered at 08:45. The strategy
refills only at prices below the forecast band of the respective product, so
it does not buy back into the ongoing spike. The refill purchases are
executed once prices have returned to their normal level. The last refill takes only part of its transaction's volume, because recovery is sized exactly to the distance from the target state of charge (meaning that we assume that we can match a transaction from our dataset partially). Netting the recovery cost against the avoided purely financial loss, the backstop is €1\,159 better off on this product and trading cycle.

Table~\ref{tab:results_main} shows the full year results for all forecasting algorithms and both trading modes. The backstop improves the profit of 13 of the 16 configurations. In the long/short mode, it improves the profit in 7 of 8 configurations, with gains between €377k and €1.81m, and it changes the risk profile fundamentally: the profit-to-risk ratio rises from a range of 0.0145 to 0.0990 to a range of 0.475 to 0.745. The mechanism only intervenes when observed prices rise beyond the distress threshold, but the volume it settles is precisely the volume that produces the tail risk.

The results also show a limit of the mechanism. Three configurations lose
profit, all of them median strategies, with the day-before Q50 strategy in
the long-only mode losing €503k. The physical settlements are not
the cause, because they only replace trades the market prices beyond the
distress threshold. The cause is the rule that halts trading once the daily
cycle budget is exhausted. The Day-Before Q50 strategy reaches its cycle budget on 360 of the 365 days, in the median already by 07:01, and opens no further positions for the rest of the day. Running the identical configuration with the continue rule instead of the halt rule earns €896k more while using the same 365 cycles.

\begin{table*}[t]
\centering
\caption{Full-year 2024 results. Fin. and phys. refer to financial and physical traded volumes, respectively.}
\label{tab:results_main}
\resizebox{\textwidth}{!}{%
\begin{tabular}{lrrrrrrrrrrrrr}
\toprule
& \multicolumn{5}{c}{Purely financial}
& \multicolumn{5}{c}{Asset-backed}
& \multicolumn{3}{c}{BESS} \\
\cmidrule(lr){2-6}\cmidrule(lr){7-11}\cmidrule(lr){12-14}
Strategy
& \shortstack{Profit\\(€k)}
& \shortstack{Profit-to-risk\\ratio}
& \shortstack{Maximum\\drawdown (€k)}
& \shortstack{Volume\\(GWh)}
& \shortstack{Profit per\\MWh (€)}
& \shortstack{Profit\\(€k)}
& \shortstack{Profit-to-risk\\ratio}
& \shortstack{Maximum\\drawdown (€k)}
& \shortstack{Volume fin.\,/\,phys.\,/\,total\\(GWh)}
& \shortstack{Profit per MWh\\fin.\,/\,phys.\,/\,total (€)}
& \shortstack{Profit gain\\(€k)}
& \shortstack{Cycles\\used}
& \shortstack{Profit per\\cycle (€)} \\
\midrule
\multicolumn{14}{l}{\textit{Long-only}} \\
\midrule
Day-before Q50 & 1,430 & 0.642 & -31.2 & 1,170 & 1.22 & 923 & 0.669 & -1.56 & 384\,/\,29.3\,/\,413 & 2.41\,/\,31.6\,/\,2.24 & -503 & 365 & -1,380 \\
Day-before Q10/90 & 881 & 0.440 & -32.1 & 460 & 1.91 & 1,080 & 0.665 & -14.6 & 216\,/\,28.2\,/\,244 & 5.02\,/\,38.3\,/\,4.44 & +201 & 352 & 570 \\
LQR Q50 & 1,480 & 0.661 & -34.4 & 1,330 & 1.12 & 1,450 & 0.891 & -12.2 & 826\,/\,27.1\,/\,853 & 1.75\,/\,53.3\,/\,1.70 & -36.9 & 338 & -109 \\
LQR Q10/90 & 694 & 0.359 & -38.2 & 246 & 2.82 & 1,160 & 0.637 & -17.0 & 188\,/\,15.5\,/\,204 & 6.17\,/\,75.0\,/\,5.70 & +468 & 193 & 2,420 \\
LSTM Q50 & 1,450 & 0.624 & -62.2 & 1,370 & 1.06 & 1,810 & 0.744 & -4.54 & 979\,/\,25.5\,/\,1,000 & 1.85\,/\,71.1\,/\,1.81 & +361 & 318 & 1,130 \\
LSTM Q10/90 & 624 & 0.349 & -75.4 & 214 & 2.91 & 1,130 & 0.599 & -8.38 & 171\,/\,10.1\,/\,181 & 6.57\,/\,112\,/\,6.21 & +502 & 126 & 4,000 \\
OrderFusion Q50 & \textbf{1,560} & \textbf{0.703} & -20.7 & 1,380 & 1.13 & 1,820 & 0.772 & -13.1 & 1,010\,/\,24.8\,/\,1,030 & 1.81\,/\,73.6\,/\,1.76 & +261 & 309 & 845 \\
OrderFusion Q10/90 & 713 & 0.500 & \textbf{-15.0} & 210 & \textbf{3.39} & 1,180 & 0.560 & -7.05 & 180\,/\,7.99\,/\,188 & 6.57\,/\,148\,/\,6.29 & +467 & 99.6 & 4,680 \\
\midrule
\multicolumn{14}{l}{\textit{Long/short}} \\
\midrule
Day-before Q50 & 703 & 0.0738 & -390 & 1,920 & 0.366 & 497 & 0.715 & -6.94 & 326\,/\,29.3\,/\,356 & 1.52\,/\,17.0\,/\,1.40 & -206 & 365 & -565 \\
Day-before Q10/90 & 307 & 0.0330 & -388 & 861 & 0.357 & 760 & 0.611 & -32.9 & 238\,/\,29.2\,/\,267 & 3.20\,/\,26.0\,/\,2.85 & +453 & 365 & 1,240 \\
LQR Q50 & 738 & 0.0833 & -358 & 2,170 & 0.340 & 1,110 & 0.745 & -6.79 & 853\,/\,29.0\,/\,882 & 1.31\,/\,38.5\,/\,1.26 & +377 & 361 & 1,040 \\
LQR Q10/90 & 233 & 0.0264 & -472 & 477 & 0.489 & 1,660 & 0.475 & -117 & 302\,/\,22.0\,/\,324 & 5.49\,/\,75.4\,/\,5.11 & +1,430 & 274 & 5,200 \\
LSTM Q50 & 686 & 0.0816 & -294 & 2,180 & 0.315 & 1,530 & 0.664 & -11.2 & 1,110\,/\,28.3\,/\,1,140 & 1.37\,/\,53.9\,/\,1.34 & +840 & 353 & 2,380 \\
LSTM Q10/90 & 119 & 0.0145 & -358 & 367 & 0.324 & 1,930 & 0.525 & -96.8 & 269\,/\,16.4\,/\,285 & 7.19\,/\,118\,/\,6.78 & \textbf{+1,810} & 204 & 8,890 \\
OrderFusion Q50 & 866 & 0.0990 & -315 & 2,190 & 0.396 & 1,670 & 0.720 & -12.8 & 1,190\,/\,28.3\,/\,1,220 & 1.40\,/\,58.9\,/\,1.36 & +801 & 353 & 2,270 \\
OrderFusion Q10/90 & 254 & 0.0324 & -362 & 363 & 0.700 & \textbf{2,060} & 0.605 & -96.6 & 299\,/\,13.4\,/\,312 & 6.89\,/\,154\,/\,6.60 & +1,810 & 167 & \textbf{10,800} \\
\midrule
\multicolumn{14}{l}{\textit{Benchmark}} \\
\midrule
Rolling intrinsic & -- & -- & -- & -- & -- & 1,770 & 0.825 & -14.6 & 224\,/\,29.3\,/\,253 & 7.91\,/\,60.4\,/\,6.99 & -- & 365 & 4,840 \\
Day-ahead auction & -- & -- & -- & -- & -- & 1,370 & \textbf{1.26} & \textbf{0} & 0\,/\,29.3\,/\,29.3 & --\,/\,46.8\,/\,46.8 & -- & 365 & 3,750 \\
\bottomrule
\end{tabular}}
\end{table*}

The comparison with the RI benchmark, which operates the identical BESS
under identical costs and cycle budget, addresses the central question of
this study. The RI benchmark earns €1.77m
and uses all 365 available cycles, earning €4\,844 per cycle. The day-ahead benchmark stays below this even with perfect foresight of the clearing prices, at €1.37m and €3\,751 per cycle, which mirrors the finding of \cite{miskiw2025continuous} that the IDC is the more valuable market for a BESS. The asset-backed OrderFusion Q10/90 strategy earns €2.06m and uses 167 cycles, which is less than half of the available budget, at €10\,798 per cycle.

\begin{figure}[t]
\centering
\includegraphics[width=\columnwidth]{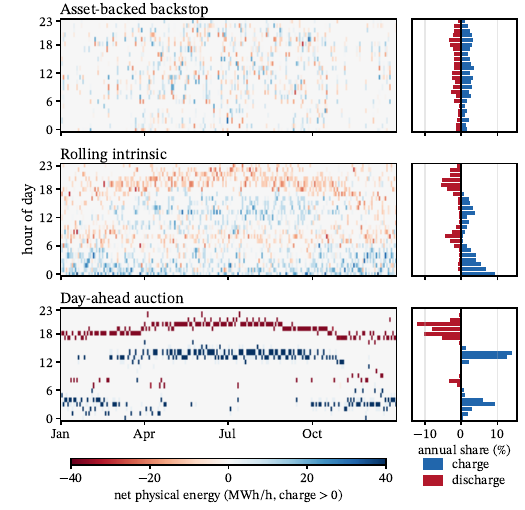}
\caption{Physical BESS usage over 2024 for the asset-backed backstop
(OrderFusion Q10/90, long/short), the rolling intrinsic benchmark and the day-ahead benchmark.}
\label{fig:BESS_usage}
\end{figure}

The number of cycles the backstop consumes also depends on the forecast. In
the long/short mode with the outer band, OrderFusion requires 167 cycles, the LSTM 204, the LQR 274, and the day-before benchmark the full budget of 365. The profit per cycle falls in exactly the same order, from €10\,798 for OrderFusion to €1\,243 for the day-before benchmark. A better forecast therefore does not only raise the
trading profit, it also reduces how much BESS capacity a trading desk
needs in order to protect the same strategy.

Fig.~\ref{fig:BESS_usage} shows when each strategy physically uses the
BESS, showing the substantial differences in time of usage. While the day-ahead strategy has a clear pattern, trading is distributed throughout the day in the RI and backstop strategies.

\textbf{Sensitivity analysis:}
\label{sec:sensitivity} Table~\ref{tab:sensitivity} varies one parameter at a time around the main configuration of the OrderFusion Q10/90 long/short strategy, showing the profit gain over the purely financial baseline as in Table~\ref{tab:results_main}.

The width factor of the distress thresholds is the parameter that matters
most. Doubling this factor from 0.5 to 1.0 makes the BESS intervene only when the
settlement price deviates from the trading band by the full forecast
bandwidth instead of half of it. This change reduces BESS usage by 59\%, from 167 to 69 equivalent full cycles, while the profit gain falls by only 6\%. The profit per cycle more than doubles, from €10\,798 to €24\,461. Reducing the width factor yields diminishing marginal returns: this raises the profit only slightly but consumes many cycles. For an operator, the width factor is therefore the main lever to free up BESS cycles at little cost to overall profit, for example to reserve the asset for ancillary services or bidding across multiple markets. The corresponding lever on the trading side is the entry band itself, which was evaluated in an additional sensitivity experiment: narrowing it from Q10/90 to Q25/75 with the OrderFusion model opens positions closer to the median and therefore far more often, which raises the purely financial profit from €254k to €413k but leaves the asset-backed strategy slightly worse off at €2.00m against €2.06m while consuming 273 instead of 167 cycles, which advocates for the choice of a wider Q10/90 band.

The daily cycle budget and the rule regarding financial trading after the daily cycle cap has been reached have a much smaller effect
on the benefits of this strategy, because it uses less than half of the available budget in
the first place. Raising the budget to two cycles per day or removing it entirely increases the profit gain by around 8\% and BESS usage by 11\%, leaving the profit per cycle almost unchanged. Continuing to trade financially after the cap instead of halting adds 0.8\% at an identical number of cycles. These parameters become mainly important for strategies that exhaust the budget, as in the day-before Q50 case.

Doubling the bid-ask spread from €1/MWh to €2/MWh, which reverses the halving
of the estimate in \cite{kuppelwieser2021liquidity} discussed in
Section~\ref{sec:casestudy}, changes the profit gain by 1.2\%. The
purely financial strategy is far more exposed to that assumption than the
BESS is: its full-year profit falls from €254k to €80k following the
same change. The value of the backstop is therefore robust to the cost assumption even where the underlying strategy is not.

The last two rows scale the trading budget and the BESS together. The purely financial strategy scales poorly with increasing trading volumes, turning a €254k profit into a loss of €390k at twice the budget and €1.96m at four times, because the settlement tail grows faster than the edge. The asset-backed strategy stays profitable, at €3.10m and €5.01m, but its return per unit of BESS wear declines, from €10\,798 per reference cycle to €9\,143 and then €8\,025.

This sheds light on one important aspect that has to be kept in mind when evaluating the results of our study. The stated trading profits are based on relatively small volumes. Increasing trading volumes and the associated BESS size would not lead to commensurate profit scaling, since there are fewer profitable trading opportunities due to liquidity constraints. In the following section, we discuss why this is an important aspect to consider, but does not harm the overall relevance of our results.

\begin{table}[t]
\centering
\caption{Sensitivity analysis for OF Q10/90 long/short strategy.}
\label{tab:sensitivity}
\resizebox{\columnwidth}{!}{%
\begin{tabular}{lrrrrrr}
\toprule
& \multicolumn{2}{c}{Profit gain (k EUR)}
& \multicolumn{2}{c}{Cycles used}
& \multicolumn{2}{c}{Profit per cycle (EUR)} \\
\cmidrule(lr){2-3}\cmidrule(lr){4-5}\cmidrule(lr){6-7}
Setting & Value & Change & Value & Change & Value & Change \\
\midrule
Main configuration & 1,806 & -- & 167 & -- & 10,798 & -- \\
Width factor 1.0 instead of 0.5 & 1,696 & \better{6.1} & 69 & \better{58.5} & 24,461 & \worse{126.5} \\
Two cycles per day instead of one & 1,948 & \worse{7.9} & 185 & \worse{10.6} & 10,532 & \better{2.5} \\
Unlimited cycles & 1,951 & \worse{8.1} & 186 & \worse{11.2} & 10,496 & \better{2.8} \\
Continue trading after the cap & 1,821 & \worse{0.8} & 167 & \same & 10,887 & \worse{0.8} \\
Bid-ask spread 2 instead of 1 EUR/MWh & 1,830 & \worse{1.3} & 167 & \better{0.2} & 10,964 & \worse{1.5} \\
Twice the trading budget and BESS & 3,488 & \worse{93.2} & 191 & \worse{14.1} & 9,143 & \better{15.3} \\
Four times the trading budget and BESS & 6,970 & \worse{286.0} & 217 & \worse{29.8} & 8,025 & \better{25.7} \\
\bottomrule
\end{tabular}}
\end{table}

\section{Discussion}
\label{sec:discussion}

Our trading strategy is a proof-of-concept for an approach to using BESS in the IDC which is new to academic research. This strategy can outperform purely financial, purely physical (day-ahead) and hybrid physical/financial (RI) approaches in absolute terms, with a modest risk trade-off and similar profit per MWh traded against the strongest performing benchmark. Not only does this key result address our research question, but we argue that this outperformance \textit{understates} the true benefits of our approach --- our strategy delivers higher revenue within the BESS cycle budget. This frees up BESS capacity for other revenue stacking opportunities. For example, the operator could combine our approach with reserving some capacity in hours the asset typically sits idle (e.g. overnight, as in Figure~\ref{fig:BESS_usage}) to bid into other markets, such as providing balancing energy or ancillary services. This additional revenue (not reported here) could materially improve the project's revenue, subject to conflicts with the overall backstop strategy, and would be a worthwhile extension of our research. Beyond this extension, we have closed all positions either financially or physically throughout, in line with German market requirements. However, the results from this strategy may change when compared against an alternative in which open positions are carried through for settlement at the imbalance price, as permitted in Great Britain or Denmark (for example).

While these results answer our primary research question, we would like to address \textit{why} our approach is able to outperform these benchmarks. Comparing the volumes traded in the purely financial and the RI strategies to the asset-backed approach, it is clear that the edge does not come solely from trading higher volumes: financial volumes and per-MWh total profits are comparable between RI and asset-backed strategies, while the purely financial strategy trades higher volumes at a lower profit. The main benefit of our asset-backed strategy (over the purely financial counterpart) is evident in comparing the maximum drawdown columns: the unit's primary role is to cap the cost of closing an adverse position,
and it is these settlement costs that seriously reduce cumulative returns (see Figure~\ref{fig:id1_modes_cumpnl}). This allows our strategy to benefit from the outsized returns available through financial trading, capturing extreme prices, while insuring against adverse market moves. Compared to the RI benchmark, we trade off a slightly more stable revenue stream for greater trading opportunities.

This framing provides an intuitive understanding for why the benefits of BESS are significantly larger for the Q10/Q90 strategies. Our quantile forecasts are produced once, and not re-evaluated conditional on the current market price when executing a trade. The market moving to (say) a Q90 level raises the conditional likelihood of observing an even more extreme price when it comes to settlement time, which we do not account for. As such, these forecasts tend to underestimate the risks of an extreme quantile-based strategy. The BESS settles these positions physically instead of at the distressed
exchange price, and the strategy halts further financial trading while the
SOC recovers. Given the cross-correlation of intraday price movements among neighbouring products, adverse moves in one period are likely to mean adverse moves in the next. Stopping trading in this period to recover SoC prevents these successive losses, while still earning income on the physical power delivery. Using the physical asset, and then recovering SoC at favourable prices, reduces the downside risk of the pure financial trading strategy. 

This naturally suggests that the benefits of our strategy are highly contingent on both the control signal for the BESS and the financial trading strategy itself: a more profitable financial strategy would see a lower marginal benefit from asset backing. In Table~\ref{tab:results_main}, the more profitable financial strategies gain less than from asset backing than their less profitable counterparts. We frame our research as a starting point for a new sub-field of BESS trading literature, in which more sophisticated financial trading and BESS control strategies for the backstop should be explored. That said, our results establish the viability of our simple trading strategy, outperforming established benchmarks from financial, physical, and hybrid trading strategies, while conserving cycles. Despite our reliance on pre-defined hyperparameters, our analysis in Table~\ref{tab:sensitivity} finds that our results are stable with respect to small changes to these values, indicating that profits from our strategy are robust. Notably, our chosen width factor ($\kappa = 0.5$) seems to represent a trade-off between rare-event coverage with poorer profit outcomes, and overspending the cycle budget.

Our analysis relies on two assumptions about the BESS and our execution strategy, which we now discuss. First, throughout we have neglected degradation costs to our unit. Total degradation costs are related to the energy throughput, and our headline result uses significantly fewer cycles than our benchmark strategies. As such, we expect that accounting for degradation costs will have little effect on the sign or magnitude of our comparative results. Second, we assume that our strategy could have executed a trade at the same price/volume as any actual historical transaction. This assumes the market was liquid enough to accommodate our order
alongside the existing transaction, which is reasonable for our small
portfolio in a market as liquid as Germany's close to delivery, applies
equally to our benchmarks, and is supported by the spread sensitivity in
Table~\ref{tab:sensitivity}. It does, however, limit the scalability of
the strategy: the volume-scaling rows show that profits do not grow
commensurately with size, and order book data are needed to assess larger
portfolios.

\section{Conclusion}
\label{sec:conclusion}

This paper examines a role for BESS on the IDC in which the asset serves as a physical backstop for a financial trading strategy. A quantile-based financial strategy opens and closes positions on the exchange, and the BESS settles physically only those positions whose financial settlement would occur at prices the forecast deems highly improbable. We backtest our strategy on 2024 German transaction data, and find that a 40~MWh BESS in this role earns €2.06m against €1.77m for an RI benchmark, and €1.37m for a perfect-foresight day-ahead benchmark, while consuming 167 instead of 365 available equivalent full cycles. We attribute the trading edge predominantly to cheaper settlement of adverse positions rather than to higher traded volume, and the freed cycles remain available for participation in further markets. 

We highlight three potential avenues for further research. First, the strategy operates on transaction data, so an implementation on order book data is required to assess execution scalability beyond the small portfolio considered here. Second, the trading thresholds, upon which the financial trading strategy is based, are fixed at cycle open and are not re-evaluated as prices move. Our results suggest that this is the main source of the distressed settlements the BESS absorbs. Conditional or time-variant thresholds could reduce their frequency and hence yield scope for future advancements in financial intraday trading research. Third, bidding of the BESS with our backstop strategy on multiple markets, for instance day-ahead, ancillary services, and imbalance settlement in certain markets should be evaluated with the spare BESS cycles.


\section*{Acknowledgment} The authors thank Peter Tankov (CREST, ENSAE Paris, Institut Polytechnique de Paris) for helpful discussions on the nature of the continuous intraday market and for valuable feedback on the manuscript. 

During the preparation of this work, the authors used Grammarly and Claude Opus 5 and Fable 5 for language refinement, grammar correction, LaTeX code generation and to support the implementation of the simulation and evaluation code. After using these tools, the authors reviewed and edited the content as needed and take full responsibility for the content of the publication.

\bibliographystyle{IEEEtran}
\bibliography{references}
\end{document}